\documentclass[%
 reprint,
 amsmath,amssymb,
 aps,
pra,superscriptaddress
]{revtex4-2}

\usepackage{bm}

\usepackage{graphicx}
\usepackage[usenames]{color}
\usepackage{physics}
\usepackage{siunitx}
\usepackage{xcolor}
\usepackage{setspace}
\usepackage{float}
\usepackage{tabularx}
\usepackage[linkcolor=blue,citecolor=blue,colorlinks=true,urlcolor=blue]{hyperref}
\usepackage{mathrsfs}
\usepackage{amsmath,esint}
\usepackage{comment}
\usepackage{lipsum}
\usepackage[toc]{appendix}

\begin{document}

\preprint{APS/123-QED}

\title{On efficiency of double ionization of a three-electron atom in moderate laser field intensities}
\author{Dmitry K. Efimov}
\email{dmitry.efimov@pwr.edu.pl}
\affiliation{Institute of Theoretical Physics, Faculty of Fundamental Problems of Technology,
Wrocław University of Science and Technology, 50-370 Wrocław, Poland}

\date{\today}

\begin{abstract}
We study double ionization in a one-dimensional model of a three-electron atom exposed to strong short laser fields. The corresponding ab initio simulations reveal a characteristic ``knee'' in the ionization yield at lower field intensities than in comparable two-electron systems. To understand the origin of this shift, a quantitative semi-analytic model for double ionization of a three-electron atom has been created. In this model, ionization is treated as a combination of different scenarios and channels. Each channel is simulated with either Quantitative Rescattering Theory (QRS) or two-electron ab initio model. Analysis of the semi-analytical results indicates that the shifted part of the ``knee'' is dominated by a recollision-excitation channel in which the rescattering electron initially escapes the parent atom from an inner orbital. This counterintuitive phenomenon is then discussed and explained.
\end{abstract}

\maketitle


\section{\label{sec:intro}Introduction}
Study of the interaction of strong laser fields with atoms and molecules has revealed a range of fascinating phenomena, from above-threshold ionization \cite{Becker2002-ga,Benda2021-bh,PhysRevA.108.053106}, non-sequential multiple ionization \cite{Pullen2017-nl,Prauzner-Bechcicki2005,Kubel2014,Becker2012-ws}, and high-harmonic generation \cite{PhysRevLett.68.3535,PRXQuantum.5.040319,Suarez2017-iv} to the production of attosecond pulses\cite{PhysRevLett.103.183901,Yan2024-fd,PhysRevResearch.6.L012020}. These advances have not only deepened our understanding of ultrafast electron dynamics, but have also led to the development of powerful experimental techniques, such as attosecond streaking and attosecond transient spectroscopy~\cite{Borrego-Varillas2022}, allowing precise measurement and control of the motion of electrons in matter on attosecond timescales.

Ionization is the most common phenomenon observed when atoms or molecules interact with a strong laser field, and the corresponding yield is one of the most important experimental sources of information about strong-field atomic dynamics. The yield of multiple ionization can serve as a tool for testing electronic correlations during atomic interaction with an external strong laser pulse \cite{Becker2012-ws,Jiang2020-ny,Wang2024-ms}. While the general shape of single ionization yield dependence on laser intensity is well described within single-active-electron theories \cite{Tong2005-kq,PhysRevA.81.033416}, one fails to describe the double ionization yield magnitudes by employing just single/independent electrons models. For the case of double ionization one usually assumes that there is always an electron that leaves the atom first, and an electron whose escape happens later in time; we will call them the first and the second electrons correspondingly. Since the laser field is AC, some portion of the first electrons may return to the parent ion afterward. It was shown that in the case of helium atoms the interaction of electrons upon recollision of the first ionized electron with the parent ion is responsible for the formation of the so-called ``knee'' in the dependence of the double ionization yield on field intensity: the rescattering electron can transfer part of its energy to the bound electron, thus facilitating the escape of the latter \cite{Lafon2001-bx,Becker2012-ws}.

Double ionization of multiple electron atoms in strong fields constitutes a much more complex phenomenon to describe because there is a larger number of interactions between particles and a much more complex electronic structure of atoms possessing three or more electrons \cite{Becker2012-ws}. Creating a descent theory for that is highly desirable not only in strong-field physics, but also in attosecond chemistry\cite{Nisoli2017-sn}. However, this task is neither simple from the perspective of numerical ab initio models nor from that of analytical models. For ab initio grid-based methods, which are the main tools used to simulate both ionization yields and momentum distributions, the numerical complexity of the model increases exponentially with the number of electrons. Currently, it is only possible to simulate the dynamics of two-electron atoms in full three-dimensional space, and only for relatively short pulses \cite{Parker2006-jv,Colgan2013-kc}. Three-dimensional quantum simulations involving three or more electrons are beyond the capabilities of modern computational methods. For such systems, only semiclassical trajectory-based calculations have been performed \cite{Peters2021-zy,Jiang2021-js,Jiang2022-qx,Katsoulis2024-vm}.  Quantum simulation of the ionization of three-electron systems on a grid has only been performed for a reduced-dimensionality model, in which each electron is confined to a chosen straight line in 3D space that passes through the atomic core \cite{Thiede2018,Ruiz05,Jiang2023-yb,Efimov2021}. The most recent analytical models for double ionization based on the Strong Field Approximation (SFA) formalism are capable of predicting yields and momentum distributions for two-electron atoms\cite{Amini2019,Hashim2024-wf}, but extending the method to systems with a larger amount of electrons seems barely feasible due to the high complexity of the corresponding analytical expansion. Semi-analytical methods based on Quantitative Rescattering Theory (QRS) are now capable of predicting momentum distributions and ionization yields of two-electron atoms with experimental precision \cite{C_D_Lin_Anh-Thu_Le_Cheng_Jin_Hui_Wei2018-md,Chen2007-ng,Chen2011-iz,Chen2021-li,Liu2022-gd}. The power of the QRS approach lies in its conceptual simplicity and modularity: once the ionization channels are determined, the QRS application is decomposed into computing the corresponding wavepackets and calculating sets of cross sections. Due to these advantages, we will use the QRS in our research.

The study of double ionization of three-electron atoms that we have performed with an ab initio atomic model with reduced dimensionality \cite{Efimov2019} has revealed two striking observations. First, double ionization of three-electron atoms cannot be simulated by a simple combination of numerical ab initio two-electron models, and the second, the double ionization ``knee'' is located within a region of smaller laser intensities in comparison to knees obtained with various two-electron ab initio models. This effect is caused by the three-electron configuration of the atomic model. Understanding the origin of the characteristic shape of double ionization knee of the three-electron model would open the road for building more effective and precise models for dynamics of many-electron atoms in strong laser fields.

In this paper, we investigate the factors that lead to the formation of such a characteristic shape of double-ionization knee and further create a semi-analytical model that predicts the shape of the discussed dependence. For achieving our goal, we provide a series of observations on the existing data and create a physical model for strong-field ionization of a three-electron atomic model, decomposing the whole phenomenon into a set of single- and two-electron processes. The latter, due to their relative simplicity, can be described with the existing ab initio models, or by the QRS model that we have already used in its simplified form for study of ionization of the three-electron totally antisymmetric model \cite{Efimov2020}.

The paper is organized as follows. In the second section, we describe both numerical and semi-analytical models that we use in the current research. It is followed by a section that contains an analysis of numerical data, a derivation of the semi-analytical model, and a discussion of its properties and predictions. The work ends with the Conclusions section. Atomic units are used throughout the paper unless specified otherwise.

\section{Atomic models}

\subsection{Three-electron ab initio model}\label{sec:3Emodel}

The \textit{ab initio} three-electron numerical model used here is identical to the model used in \cite{Efimov2019}. In the model, the position space contains three dimensions, allocating one dimension per each electron. The motion of electrons is thus one-dimensional and can be interpreted as if a 3D electron would move along a straight line. The lines corresponding to each of the electrons are said to constitute the angle $\pi/6$ between any two and the angle $\arctan (\sqrt{2/3})$ with the direction of the linearly polarized laser field. The interelectron and electron-core interactions are modeled with soft-core potentials that account for the model geometry. The corresponding Hamiltonian reads:

\begin{multline}
H=\sum_{i=1}^3 \frac{p_i^2}{2}  -\sum_{i=1}^3\left(\frac{3}{\sqrt{r_i^2+\epsilon^2}}+\sqrt{\frac{2}{3}} F(t) r_i\right) \\
+\sum_{i, j=1, i<j}^3 \frac{1}{\sqrt{\left(r_i-r_j\right)^2+r_i r_j+\epsilon^2}},
\label{Ham1}
\end{multline}

\noindent with $r_i$ and $p_i$ being the coordinate and momentum of $i$' the electron, respectively, $\epsilon = \sqrt{0.83}$ is a parameter that softens the Coulomb singularity, and the field $F(t)=-\partial A / \partial t$ depends on the vector potential.

\begin{equation}
A(t)=\frac{F_0}{\omega_0} \sin ^2\left(\frac{\pi t}{T_p}\right) \sin \left(\omega_0 t\right), \quad 0<t<T_p
\end{equation}

Here $F_0$ denotes the field amplitude, the field frequency value $\omega_0=0.06$a.u. is taken to be in agreement with our previous studies; the pulse length $T_p=2 \pi n_c / \omega_0$ is taken to be a multiple of the number of cycles $n_c=5$.

The numerical simulations have been performed on a numerical grid covering 400 a.u. of position space along each direction with a grid step equal to $100/512$ a.u. The initial state used for the time-dependent Schrodinger equation (TDSE) has been computed for the introduced Hamiltonian (\ref{Ham1}) using the imaginary-time method. The ground state is chosen to be partially antisymmetric, that is, the wavefunction is symmetric in respect to permutation of positions of one pair of electrons, but not symmetric in respect to permutations within the remaining two pair of electrons. Physically, such a configuration corresponds to a system of electrons with different spins: the two electrons with the same spin that we call ``U'', and one electron with a different spin that we call ``D'' \cite{Thiede2018,Efimov2019,Efimov2020,Prauzner-Bechcicki2021}.

\begin{figure}
	 \includegraphics[width=1.0\linewidth]{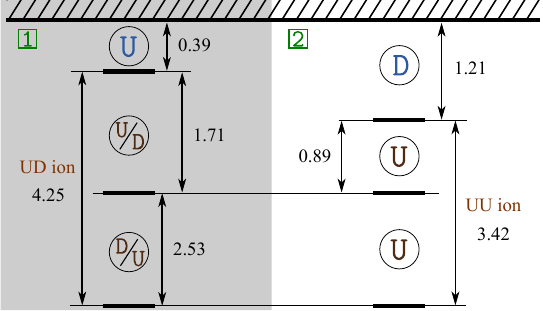}
	 \caption{Level schemes for different ionization scenarios (marked
with numbers in the green squares). Letters in circles denote spin of the electron ionized in a particular event. Their relative position from the top to the bottom indicates the order at which electrons leave atom/ion. Ionization events are visualized from the top to the bottom; the first ionized electrons are colored in blue, the type of remaining single ion is marked in brown. The notation (U/D) vs (D/U) for the second and third ionization events refers to situation when ionization of U electron is followed by that of D electron or vice versa. Numbers reflect the ionization potential values that were calculated for the ab-initio three-electron model (\ref{Ham1}) with parameters' values specified in sec. \ref{sec:3Emodel}. }
   \label{fig0}
\end{figure}

A method for computing the yields of particular ionization channels is described in the following. The position space is divided into areas that we say correspond to different states of atom, with respect to the number and spin (U/D) of ionized electrons (see \cite{Thiede2018} for how this division is done). Such states are the neutral one (N), singly ionized with free U- or D-electron (SI--U/SI--D), doubly ionized with pairs \{U,U\} or \{D,U\} of free electrons (DI--UU/DI--DU), and triply ionized states. We say that transitions between the above states define ionization channels. We first define them formally, and then discuss their straightforward physical meaning later in the paragraph. These ionization channels include single ionization with respect to the U-electron, denoted as SI(U): N $\to$ SI–-U, and similarly, single ionization with respect to the D-electron: SI(D): N $\to$ SI--D. Time-delayed double ionization can occur via several pathways. The first is TDI(0-U-U): N $\to$ SI--U $\to$ DI--UU. The second is TDI(0-D-U), which proceeds as N $\to$ SI--D $\to$ DI--DU; and the third TDI(0-U-D) implies N $\to$ SI--U $\to$ DI--DU. Recollision-induced double ionization is also possible, either as RII(0-UU): N $\to$ DI--UU or as RII(0-DU): N $\to$ DI--DU \cite{Efimov2019}. The described ionization channels can be grouped into two ionization scenarios illustrated in Fig. \ref{fig0}: when the U electron is ionized first and when the D electron is ionized first. From the definition, the naive physical interpretation of the double ionization channels follows: for TDI, electrons escape atom one-by-one: a neutral atom becomes the first single ion and afterwards evolves into a double ion, ionization through Recollision, excitation with subsequent ionization (RESI) or sequential ionization are not distinguished.
For RII, singly ionized state is eliminated; this may happen in the case of direct ionization, that is, the energy of recolliding electron is larger than the ionization potential of the bound electron, or, for instance, when the external field is strong and short enough to sequentially ionize both electrons within very short time around local maximum of laser field so that both electrons escape the atom in the same direction. An in-depth discussion of channel interpretation is presented in Sect.~\ref{sec:DnR}. Throughout the paper, we reserve the notions of ``RII'' and ``TDI'' for description of the results of the ab initio model, while ``Direct ionization'' and ``RESI'' are used for denoting physical processes that provide to ionization, in the context, for example, of the semi-analytical model.

\subsection{Two-electron ab initio model}\label{sec:2Emodel}

The two-electron \textit{ab initio} model \cite{Prauzner-Bechcicki2007,Prauzner-Bechcicki2008,Efimov2018} is used here to simulate the ionization of a single ion of the model atom. Removing one electron from the three-electron atom results in a two-dimensional model, which preserves the angles that are formed by axes of the electron's motion and the direction of the electric field \cite{Efimov2019}. The nuclear charge is set accordingly to 3. The Hamiltonian of the problem reads as follows:

\begin{multline}
H=  \sum_{i=1}^2\left(\frac{p_i^2}{2}-\frac{3}{\sqrt{r_i^2+\epsilon^2}}+\sqrt{\frac{2}{3}} F(t) r_i\right) \\
 +\frac{1}{\sqrt{\left(r_1-r_2\right)^2+r_1 r_2+\epsilon^2}}.
\end{multline}

The algorithm for numerical solution is similar to the one used in three-electron model, the grid length along one dimension and the grid step are the same as in three-electron model. The wavefunction considered here is spatially antisymmetric in case it reproduces a system of two electrons with identical spin (UU), and spatially symmetric, provided it relates to a system with two different electrons (UD). The single ionization yield is computed as flux of the wavefunction over the border between neutral (N) and singly ionized (SI) regions integrated over the time of evolution.



\subsection{SFA2 model for single ion}\label{sec:SFA}

The wavepacket $W(p)$, to be defined later, can be conveniently computed with the second-order strong-field approximation (SFA2) \cite{Chen2007-ng}, since it is derived from the part of the wavefunction that actually rescatters on the parent ion during evolution in the strong field. The corresponding procedures are similar for the cases of wavepackets formed by first-ionized U- and D- electrons. We first describe the method for the latter case, and then we provide a discussion of the U-electron wavepacket calculation.  In order to get a wavepacket, one considers a single-electron problem, where initial bound state of the electron is computed from the three-electron ground state used in ab initio simulations of the time-dependent Schrodinger equation. The wavefunction, due to partial antisymmetry\cite{Efimov2023}, can be factorized as

\begin{equation}
  \Psi(r_1,r_2,r_3) = \Psi_1(r_1,r_2)\phi_0(r_3),
\end{equation}
\noindent so that $\phi_0(r_3)$ can be extracted from $\Psi(r_1,r_2,r_3)$ by fixing $r_1$ and $r_2$ at arbitrary values. The procedure ended with the shape $\phi_0 (r) \sim e^{-\beta r}$, where $\beta = 1.50$.


Within the frame of single-electron QRS, the wavepacket is defined as a ratio between the final momentum distribution $I(p)\equiv |f_2(p)|^2$ and the differential elastic scattering cross section $\sigma(p_r)$, where $f_2(p)$ is a probability amplitude of finding a free electron with momentum $p$ after rescattering at time $t\to \infty$ \cite{Chen2007-ng,Micheau2009-dg}:

 \begin{equation}
   W(p_r) = \cfrac{I(p)}{\sigma(p_r)}.
   \label{wp1}
 \end{equation}

\noindent Here, $p$ and $p_r$, both defined on the real axis, denote the 1D momentum of the electron after the end of the laser pulse and at the moment of recollision, respectively. The formal dependence of $p=-A_r \pm p_r$ on $p_r$ and on the vector potential of the laser field at the moment of electron recollision $A_r$ can be substituted with use of the standard approximation $|p_r| = 1.26 |A_r|$ \cite{Chen2007-ng,Chen2018-yr,C_D_Lin_Anh-Thu_Le_Cheng_Jin_Hui_Wei2018-md,Liu2022-gd} to $p=\{ 0.2 p_r, -1.79 p_r \}$. These two values correspond to forward and backscattering correspondingly. Following the standard approach \cite{C_D_Lin_Anh-Thu_Le_Cheng_Jin_Hui_Wei2018-md}, we further take into account only backscattering, since it dominates in the higher-energy region of the electronic spectrum, which in turn dominates in ionic excitation and direct ionization upon recollision. With these approximations, the expression for the wavepacket reads:

\begin{equation}
  W(p_r) = \cfrac{I(-1.79 p_r)}{\sigma(p_r)}.
  \label{wp2}
\end{equation}

The elastic scattering cross section is computed within the first Born approximation \cite{Chen2007-ng}. We choose the scattering potential for the one-electron model as $V_{\text{elastic}} (r) = - e^{-\alpha |r|}$, $\alpha = 1.55$. The corresponding cross section reads then:

\begin{multline}
  \sigma(p_r) = |V(q)|^2, \quad \\ V(q) = -\cfrac{1}{4\pi} \int_{-\infty}^{+\infty} e^{-iqr} V_{\text{elastic}} (r)\, dr ,
  \label{sc1E1}
\end{multline}

\noindent where $q$ is the change in momentum due to scattering; for backscattering in 1D it is equal to $-2p_r$. Substituting the shape of the potential into equation (\ref{sc1E1}) and integration gives

\begin{equation}
  \sigma(p_r) = \cfrac{1}{8\pi^3} \left| \cfrac{\alpha}{\alpha^2 + 4p_r^2} \right|^2.
  \label{sc1E2}
\end{equation}

For computing $I(p)$, we generally follow the standard SFA2 routine \cite{Lewenstein1994-gs,Amini2019,Chen2007-ng}, adjusting it to the single-dimensional model. The amplitude $f_2(p)$ is computed as

\begin{equation}
\begin{aligned}
f_2(p)= & -\int_{-\infty}^{\infty} d t \int_{-\infty}^t d t^{\prime} \int_{-\infty}^{\infty} d k\left\langle\chi_{p}(t)\right| V_{\text{elastic}}\left|\chi_{k}(t)\right\rangle \\
& \times\left\langle\chi_{k}\left(t^{\prime}\right)\right| H_i\left(t^{\prime}\right)\left|\phi_0\left(t^{\prime}\right)\right\rangle,
\end{aligned}
\end{equation}

\noindent where $H_i(t) = r\mathscr{E}(t)$ stands for the part of the Hamiltonian responsible for the interaction of an electron with an external electric field $F(t)=(2/3)^{-1/2}\mathscr{E}(t)$, $\phi_0(r,t) = \phi_0(r)\exp(iI_p t)$ with $I_p = 1.21$ denoting the ionization potential and $\left|\chi_{p}(t)\right\rangle$ denotes a Volkov state that is expressed in position representation as

\begin{equation}
\left|\chi_{p}(t)\right\rangle = \cfrac{1}{\sqrt{2\pi}}\exp(ir[p+A(t)])\exp(-iS_p(t)),
\end{equation}

\noindent with $A(t)$ being the vector potential of the external field and the action computed as

\begin{equation}
S_p(t)=\cfrac{1}{2} \int_{-\infty}^{t} d t^{\prime} (p+A(t'))^2.
\end{equation}

\noindent The variable $k$ represents the momentum of the electron between its first escape from the atom at time $t'$ and its rescattering with the parent ion at time $t$. Integration over space coordinates provides the following expression for $f_2(p)$ with precision up to a constant factor:

\begin{equation}
\begin{aligned}
f_2(p)= & \int_{-\infty}^{\infty} d t \int_{-\infty}^t d t^{\prime} \int_{-\infty}^{\infty} d k \, \exp(iS_k(t')) \\
& \times \exp(-i[S_k(t)-S_p(t)]) \exp(iI_pt') \\
& \times \cfrac{\mathscr{E}(t')\cdot (k+A(t'))}{(\alpha^2 +[k-p]^2)(\beta^2 +[k+A(t')]^2)^2}.
\end{aligned}
\label{ghy2}
\end{equation}

The resulting three-fold integral is further simplified by using the saddle-point method. In particular, we evaluate one of the three equations for the saddle points as follows:

\begin{equation}
  k\cdot(t-t') = -\int_{t'}^{t} dt'' \, A(t''),
\end{equation}

\noindent which defines the saddle point $k_0(t,t')$ with respect to the variable $k$, accounting for which reduces eq. (\ref{ghy2}) to

\begin{multline}
f_2(p)=  \int_{-\infty}^{\infty} d t \int_{-\infty}^t d t^{\prime}  \, (|t-t'|+\epsilon)^{-1/2} \exp(iS_{k_0(t,t')}(t')) \\
 \times \exp(-i[S_{k_0(t,t')}(t)-S_p(t)]) \exp(iI_pt') \\
 \times \cfrac{\mathscr{E}(t')\cdot [{k_0(t,t')}+A(t')]}{(\alpha^2 +[{k_0(t,t')}-p]^2)(\beta^2 +[{k_0(t,t')}+A(t')]^2)^2},
\label{ghy3}
\end{multline}

\noindent The integral (\ref{ghy3}) is computed numerically. The parameter $\epsilon = 10^{-10}$ is introduced to remove the corresponding singularity of the integrated function.

The computed amplitude $f_2(p)$ is further normalized by juxtaposing the ionization yield $\int I(p)\, dp$ calculated with SFA with the corresponding yield obtained with the three-electron ab initio model. The numerical curve for the D-electron cannot be used as a reference along the whole interval of field intensities since it depicts the final share of singly ionized states and not the probability of laser-assisted ionization. Because of that, we calibrate the SFA values by demanding that the D-electron ionization yields for both methods should be the same for $F=0.07$, where the double ionization outcome is small compared to single ionization (see Fig.~\ref{fig1}), and thus the ionization yield corresponds to the tunneling yield. The tunneling yield of the D-electron is connected with the amplitude as $P_{D_{tunn}} \simeq 2\int I(p)\, dp$, since one can safely assume that approximately half of the ionized electrons return to the parent ion due to restrictions on electronic motion imposed by single-dimensional geometry.

Computing wavepackets for the case of ionization of a U electron is very similar to the procedure described above. One needs to take different ionization potential $I_p = 0.39$ a.u. and initial wavefunction. In contrast to the D-electron, the initial wavefunction of the valence U electron $\phi_1(r)$ cannot be immediately extracted from the three-electron ground state due to their numerically complex relation \cite{Efimov2023}, which in the Hartree-Fock approximation would read: $\Psi_1 =  \phi_1(r_1) \phi_2(r_2) - \phi_1(r_2) \phi_2(r_1)$, $\phi_2(r)$ stands for the lower U-electron orbital. But because $\phi_1(r)$ is the upper orbital, one can approximate it with a Dyson orbital \cite{Murray2011-of}: $\phi_1(r_1) \simeq \int \Psi(r_1,r_2,r_3)  \Psi_2^*(r_2,r_3)\, dr_2\, dr_3$, here $\Psi_2(r_2,r_3)$ represents the ground state of a single UD ion. After fitting $\phi_1 (r) \sim e^{-\beta_U r}$ one gets $\beta_U = 0.75$.
Normalization for U-wavepackets is slightly different than in the case of D-electron. Since the ionization yield of the U-electron is saturated at values around 1 across the entire interval of field intensities of interest, we set $P_{D_{tunn}}(F) = 1$ and perform the normalization procedure independently for all values of $F$.



\subsection{Computation of excitation an ionization cross sections for single ion}\label{sec:CS}

For computing excitation cross sections for the model of a single UU ion, we follow the approach developed in \cite{Efimov2020}. 
In a single-dimensional model the energy conservation for a system of ion plus a colliding electron is expressed as

\begin{equation}
  (E_n-E_0) - (p^2/2 - [p-q]^2/2)=0,
\label{eq34}
\end{equation}

\noindent where $p$ is the projectile momentum before recollision, it changes by $q$ due to recollision. The energy of the ion changes at the same time from $E_0$ to $E_n$ that correspond to the initial ground ionic state $|0\rangle$ and the final excited state $|n\rangle$. The corresponding differential cross section is

\begin{multline}
  \cfrac{d\sigma_n}{d p} = \int\limits_{0}^{p} \cfrac{d\sigma_n}{d q}\, \delta \left( (E_n-E_0) - \left(\cfrac{p^2}{2} - \cfrac{(p-q)^2}{2}\right) \right) d\, q = \\ \cfrac{d\sigma_n}{d q}\bigg|_{q=q_1} + \cfrac{d\sigma_n}{d q}\bigg|_{q=q_2},
  \label{eq35}
\end{multline}

\noindent where $q_1$ and $q_2$ are real roots of equation (\ref{eq34}), if they do not exist, the value of eq. (\ref{eq35}) is 0. The expression for $d\sigma_n/d q$ is \cite{Efimov2020}:

\begin{equation}
  \begin{split}
  &\cfrac{d\sigma_n}{d q} =  \cfrac{q|c_0|^2}{8\pi p^2} \left|
\left\langle  n \left|  \mathbf{D}(q,\{ r_a \})   \right| 0 \right\rangle \right|^2,\\
&\mathbf{D}(q,\{ r_a \}) = \sum_a K_0 \left( q \sqrt{\frac{3}{4}r_a^2+\epsilon^2}\right) r_a ,
\end{split}
\label{parrot}
\end{equation}

\noindent where we assume that the ground state of the ion may be partially depleted and its population is $|c_0|^2 \in [0,1]$; $r_a$ stands for position of a valence bound electron, the summation goes over the two bound electrons in the model, and $K_0$ denotes a modified Bessel function of the second kind. Computing of eq. (\ref{parrot}) has been performed numerically for a range of values of $q$, given precomputed eigenstates of a single UU ion. Generally speaking, the population $|c_0|^2$ varies during strong field evolution, and thus its precise evaluation is complicated; it depends on the rates of laser-assisted ionization of the single ion, as well as on the intensity of recollision-induced ionization and excitation. We estimate its value for equation (\ref{parrot}) as the ionization yield of the UU ion for the given magnitude of the laser field since for relatively short laser pulses the laser-induced ionization of an ion happens mostly before recollision of the first ion. 

Calculating a direct recollisional ionization cross section is similar to computing excitation cross sections. We approximate the final states of the two-electron system after interaction with the rescattering electron as $|m_d\rangle |k\rangle$, where $|m_d\rangle $ is a triple ion bound state, described by a one-dimensional wavefunction that we compute numerically with imaginary time propagation method, and $|k\rangle = (2\pi)^{-1/2}\exp(-ikr)$ stands for an unperturbed single electron free state with momentum $k$. 
For obtaining ionization cross section, summation over all $m_d$ and integration over $k$ is necessary:

\begin{multline}
  \cfrac{d\sigma_{\text{ion}}}{d q} =  \cfrac{q|c_0|^2}{8\pi p^2}  \sum_{m_d} \int dk \,  \left| \left\langle  m_d | \langle  k |  \mathbf{D}(q,\{ r_a \})   | 0 \right\rangle \right|^2,
\label{parrot2}
\end{multline}

\noindent and $q$ is now defined from the modified equation (\ref{eq34}):

\begin{equation}
  (E_{m_d} + k^2/2 -E_0) - (p^2/2 - [p-q]^2/2)=0.
\label{eq345}
\end{equation}

\begin{figure}[t] 
	 \includegraphics[width=1.0\linewidth]{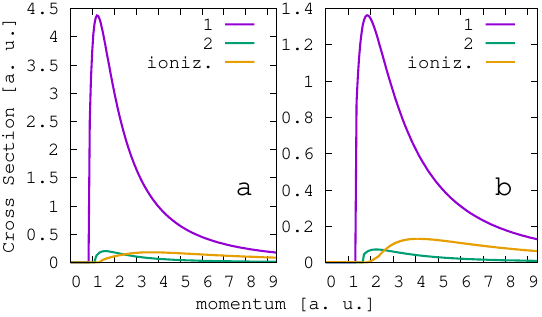}
	 \caption{Differential direct ionization (ocher curve) and excitation cross section for UU (a) and UD (b) models in dependence on momentum of recolliding electron. Excitation cross sections are depicted for the two lowest lying excited states (denoted for each model as 1 and 2) to which transitions (\ref{parrot}) from ground state are allowed. $|c_0|^2$ is set to 1 for the both plots. The excitation energies for states 1,2 of UU potential are correspondingly 0.37 and 0.66 a.u., for UD ion they are 0.9 and 1.4 a.u. }
   \label{fig_cross}
\end{figure}

The expressions (\ref{eq34}-\ref{eq345}) are valid for calculation of excitation and ionization cross sections of UD ion as well. UD ion has a different set of single-ion eigenstates $\{|n\rangle \}$,  $|0\rangle$ and the corresponding eigenenergies. The cross sections for both models are visualized in Fig.~\ref{fig_cross}

\section{Discussion and Results \label{sec:DnR}}

\subsection{The problem of early knee}

\begin{figure}
	 \includegraphics[width=1.0\linewidth]{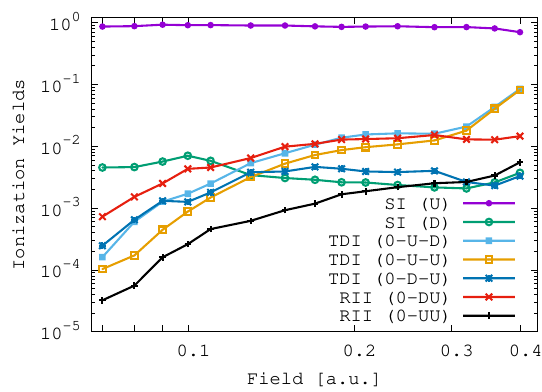}
	 \caption{Single and Double Ionization yields after ab initio three-electron model (\ref{Ham1}) in dependence on laser field amplitude for a set of ionization channels. The values of the model's parameters as well as description of ionization channels are provided in sec. \ref{sec:3Emodel}. }
   \label{fig1}
\end{figure}

To understand the origin of the double-ionization yield amplification for small fields, that is, $F<0.15$ a.u., we first discuss the results of the numerical simulation of double ionization for the three-electron model. Since a detailed description has already been provided in \cite{Efimov2019}, we restrict ourselves to the key features that are crucial for the present discussion.  In Fig. \ref{fig1}, the dominant double ionization channel for small fields is RII that involves U and D electrons. It is followed by the TDI channels, among which the most intense are the two involving U and D electrons; together, they are of comparable magnitude to the RII (0-UD) channel. The TDI(0-U-U) channel is significantly lower in magnitude than the described above, especially for the very low field intensities; the RII(0-UU) channel is about one order of magnitude smaller than the corresponding TDI channel.

The dominance of the RII channel for lower fields looks rather unintuitive, since despite the fact that direct ionization of the UD ion is energetically allowed for the range of fields depicted in Fig.~\ref{fig1}, only a small portion of returning electrons possess sufficient energy for implementation of direct ionization, and thus such channel should be statistically unimportant. This makes one suspect that the numerically defined RII channel might account for not only direct ionization, but also RESI. Consequently, rescattering-induced excitation should play a key role in the formation of the ``early knee''. However, one has yet to find out, in which of the two scenarios collisional excitation with subsequent ionization is most intense; we will pay special attention to it in Sec.~\ref{sec:recoll}.  Laser-induced ionization and further rescattering of the D electron have a chance to significantly impact double ionization for small fields, since the UU single ion, with which the D electron collides, has about twice smaller ionization potential than the UD ion with which the U electron collides (see Fig.~\ref{fig0}). At the same time, the laser-assisted ionization rate for the D electron is much smaller than that of the U electron, thus reducing the relative RESI yield of the first-D electron scenario. We shall discuss the problem in the following subsection.

\subsection{Enhanced ionization rates from an inner orbital}

The curve of SI (U), which corresponds to ionization of the upper orbital (see Fig.~\ref{fig0}), is unsurprisingly saturated near 1 for the whole range of field intensities in Fig.~\ref{fig1} due to a low ionization potential of 0.39 a.u. It is much more interesting that the values of SI(D) in Fig.~\ref{fig1} are notably high. The rough application of ADK \cite{ammosov1986tunnel,C_D_Lin_Anh-Thu_Le_Cheng_Jin_Hui_Wei2018-md} formula for $F=0.07$ a.u. gives a ratio of ionization yields of hydrogen-like atoms with ionization potentials of 1.21 a.u. and 0.39 a.u. of the order of $10^{-14}$. At the same time, the ratio of single ionization yields the first-U and first-D electron scenarios of the three-electron ab initio model, which have the same ionization potentials, is of the order of $10^{-2}$. The difference is impressive.

In order to understand such a considerable population of SI(D) states, one should account for electronic interaction during ionization, which has been approached with the Hartree-Fock (HF) formalism in \cite{Amusia2009-py}, the analytical R-matrix theory in \cite{Torlina2012-yh} and generalized in \cite{Kozlov2013-uq}. In the most intuitive HF picture, the upper orbital affects the inner one in such a way that the latter is stretched out in space, which considerably eases the escape of an electron from the inner orbital. This explains the relatively high values of SI(D) of the three-electron numerical model, since this model fully accounts for the electronic correlation.


The single ionization yield of the D electron is found to possess a rather weak dependence on the laser field amplitude (see Fig.~\ref{fig1}). Since SI(D) never reaches saturation at 1, this behavior is explained by competing with the other ionization channels. In particular, there is an interplay between the two ionization scenarios: the probability of tunnel ionization of U-electron is much larger than that of D-electron, thus during strong-field evolution, U electron escapes the parent atom fast \cite{Thiede2018} (see Fig. \ref{fig1}), severely depleting bound neutral states that could have served an initial state for D-electron ionization. 
Additionally, SI(D) competes with both sequential and recollisional double ionization that occurs after laser-assisted escape of the D electron. In the first-D scenario the second ionization potential is 1.36 times smaller than the first ionization potential, which makes sequential ionization of the second electron much more probable than the ionization of the first D electron. Such sequential ionization modeled as laser-assisted single ionization of a single UU ion experiences saturation-like behaviour around $F=0.17$ a.u. (see Fig.~\ref{fig2}) which essentially means that escape of the first D electron would almost certainly be followed by laser-assisted escape of the second (U) electron. This case will be discussed in detail in the next subsection. Both the enhanced magnitude of SI(D) yield and its limited growth are of key importance for subsequent discussion of double-ionization channels.



\subsection{Double ionization channels}



RII / TDI classification of ionization channels performed on geometric / spatial ground is standard for interpreting results of grid-based numerical models (see the corresponding description in Subsection \ref{sec:3Emodel}). Alternatively, one can introduce division based on particular types of physical processes which is more suitable for the semi-analytic tools that we aim to develop. Double ionization is conventionally divided into sequential (SDI) and non-sequential (NSDI) channels. In the first case, electrons are said to leave the atom/ion independently, and in the second case, interaction between them, when electron rescattering on the parent ion, is said to play a crucial role in ionization.

\begin{figure}[t] 
	 \includegraphics[width=1.0\linewidth]{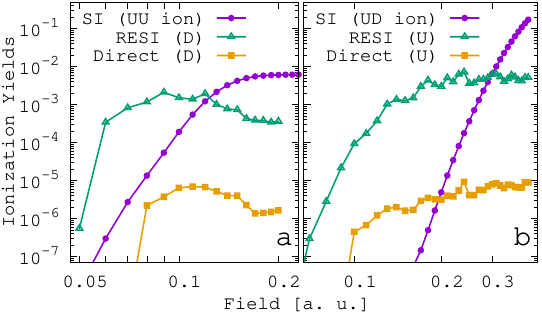}
	 \caption{Ionization Yields in dependence on laser field amplitude. Violet circles correspond to single ionization yield obtained with \textit{ab initio} two-electron model (sec. \ref{sec:2Emodel}) corresponding to UU (a) and UD (b) configurations; in (a) the corresponding yield is multiplied by a factor of $10^{-2}$. Green triangles and ocher squares correspond to yield of RESI and Direct processes after QRS calculations (sec. \ref{sec:recoll}) for the first ionized D electron (a) and U electron (b). }
   \label{fig2}
\end{figure}

For the first-D electron scenario, the notion of SDI should be refined. Indeed, a standard picture for the phenomenon implies that the two electrons escape atom one-by-one, that is, the second electron escapes the ion presumably during one of the following field maxima. In case if the second ionization potential is actually considerably smaller than the first ionization potential, the second electron may ionize during the same field maxima when the first electron escaped. In such a case, it is not clear if the escape of electrons is actually sequential, but it can be interpreted as non-recollisional. One can check its role by simulating the ionization of a single UU ion in a laser field with the same parameters that were used in three-electron simulations. The corresponding computed yield tends to 1 for $F>0.17$ a.u. (The corresponding values multiplied by $10^{-2}$ are plotted as a violet curve in Fig.~\ref{fig2}a). The corresponding yield of sequential ionization is the yield of single ionization of a UU ion normalized to the tunneling yield of the D electron. Since computing the absolute magnitude of the later quantity is a challenging task, in the simplest approach we take the maximum value $10^{-2}$ of the D-electron ionization yield from ab initio data (Fig. \ref{fig1}) as a constant factor for the SDI yield normalization in the D-electron scenario. For the case of UD ion such a factor can be taken to 1 since ionization of first U-electron is almost saturated everywhere in the plot range. The resulting double-ionization curves in Fig. \ref{fig2} (a) and (b) saturate around $F=0.17$ a.u. and for $F>0.40$ correspondingly and grow sharply for smaller values of field; thus the sequential channel does not cause the effect of ``early knee''.

\begin{figure}[t] 
	 \includegraphics[width=1.0\linewidth]{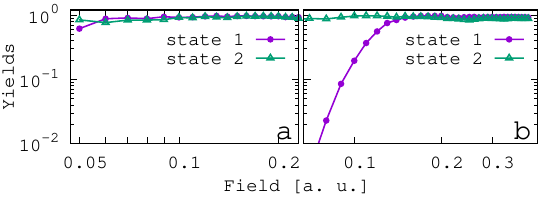}
	 \caption{Yields of single ionization of UU (a) and UD (b) single ions initially excited to various states in dependence on laser field amplitude. The two lowest states to which transitions (\ref{parrot}) from ground state are allowed are chosen.  }
   \label{fig3}
\end{figure}

The NSDI channel implies recollision of the first electron happening some time after its escape and inducing electronic dynamics that leads to ejection of the second electron. We suspect that this channel is responsible for ``early knee'' because SDI is not. NSDI can be conveniently split into direct ionization, when the recolliding electron transfers enough energy to the another electron to ionize it, while remaining in the free state itself; and RESI (recollision, excitation with subsequent ionization) which implies that the bound electron is just excited as a result of the first electron recollision, and further ionization of the resulting ionic state is provided by the laser field. For the first-D scenario, the first electron-tunneling yield has been shown to be relatively high. Thus, the success of NSDI in modest fields depends on (i) whether relatively small energies the first electron brings to the ion upon recollision are capable of providing efficient excitation or direct ionization; and in case of excitation, (ii) whether the external field can efficiently ionize the corresponding ionic states. 
The latter question can be immediately answered by simulating the ionization of excited states taken as initial ones; the results for the UU-ion are shown in Fig. \ref{fig3} and clearly mark a possibility of efficient laser-induced ionization of excited states -- saturation of the ionization yield of the first excite state happens already at $F=0.06$ a.u. At the same time, the efficiency of ionic excitation can be estimated with a quantitative model only.

We propose that the ``early knee'' can be explained with a recollisional model: the D-electron is ionized first, then it rescatters to the remaining UU-ion, exciting the latter to one of its excited states that are then tunnel-ionized by laser field.
We are going to check this hypotheses by introducing a quantitative model that combines the QRS method with providing ab initio calculations.


\subsection{Semi-analytical model \label{sec:recoll}}

For each type of single ion (UU/UD), there are three double ionization channels that can be accounted for in the model: sequential ionization, RESI, and direct ionization. The sum of the corresponding partial yields gives the total yield of double ionization:

\begin{equation}
P_{\text{tot}} = P_{\text{SDI}} + P_{\text{RESI}} + P_{\text{dir}}.
\label{wer}
\end{equation}

\noindent We compute all the mentioned yields independently. $P_{\text{SDI}}$ can be obtained with the use of a two-electron numerical model (sec. \ref{sec:2Emodel}). The corresponding ab initio simulation allows one to obtain the yield of a single ionization of the particular single ion of choice $P_{\text{SI of ion}}$. The departure of the first electron turns the atom to a single ion, which can be further ionized by a laser field. Thus, the SDI yield can be computed as a product of the corresponding yields of the mentioned partial processes:

\begin{equation}
P_{\text{SDI}} =  P_{\text{first el}}  P_{\text{SI of ion}},
\label{SDI}
\end{equation}

\noindent where both partial yields are normalized to 1. The laser-assisted ionization yield values of the first electron $P_{\text{first el.}}$ can be obtained from the data after three-electron ab initio calculations. In case of first U electron and UD ion, it can be well approximated by 1, since the SI(U) is saturated everywhere (Fig.~\ref{fig1}). For the first D electron and the UU ion, the direct extraction of $P_{\text{first el.}}$ is not possible since the numerically obtained SI(D) does not correspond directly to the laser-induced yield of the D electron due to the comparable magnitude of double ionization involving this electron. In the first approximation we set it equal to $10^{-2}$ after the maximum value of SI(D) in Fig. \ref{fig1}. The corresponding SDI yields are depicted as violet lines in Fig. \ref{fig2}.

The double ionization yield due to RESI can be described as \cite{Efimov2020}:

\begin{equation}
P_{\text{RESI}} \simeq \sum_{n} Y_{n}^{exci}\int Y_{n}^{tunn}(t)\, dt
\label{whale}
\end{equation}

\noindent where $Y_{n}^{exci}$ stands for the rate of recollisional excitation of a single UU ion from its ground state to the $n$th excited state, and $Y_{n}^{tunn}(t)$ denotes the tunneling rate from the latter. The integration of $Y_{n}^{tunn}$ should be in principle performed for the time interval starting at the moment of excitation, but it is safe to integrate over the entire period of laser-atom interaction \cite{Thiede2018}. The values of the integral of $Y_{n}^{tunn}$ are obtained as single ionization yield (see Fig. \ref{fig3}) in the numerical two-electron model for the UU / UD ion with the $n$th eigenstate of the ion taken as an initial wavefunction. As one can see, the ionization from the excited states is almost completely saturated, thus the efficiency of RESI is defined by rates $Y_{n}^{exci}$ of excitation from the ground state to higher levels.

The ionic excitation rate $Y_{n}^{exci}$ is in turn defined by multiplication of the differential excitation cross section ${d\sigma_n}/{dp}$ and the wavepacket $W(p)$ \cite{Efimov2020}:

\begin{equation}
 Y_{n}^{exci} =  \int \frac{d\sigma_n}{dp} W(p) \, dp  .
\label{whale2}
\end{equation}

\noindent We compute $W(p)$ for each amplitude of laser field with SFA modified specifically for a 1D system, as described in sec. \ref{sec:SFA}. Excitation cross sections are computed numerically from the ion's eigenstates as described in sec. \ref{sec:CS}. Normalization of recollision-based ionization yields is achieved through proper normalization of the wavepackets.

The corresponding RESI yields obtained with equation (\ref{whale}) are represented as green triangles in Fig.~\ref{fig2}. For the UD ion, the corresponding curve clearly saturates around $F=0.20$ a.u., which clearly corresponds to saturation of excitation under the condition of constant flux of electrons -- the rate of first U electron ionization is nearly constant over the whole range of field amplitude values. At the same time, the UD single ionization channel is not intense enough to deplete the ionic ground state even for the highest field values, thus it does not impact RESI efficiency. In contrast, single ionization of the UU ion saturates already around $F=0.15$ a.u. thus significantly depleting the UU ion's ground state and decreasing the magnitude of the corresponding excitation cross sections (\ref{parrot}). This results in a decrease of the RESI yield since around $F=0.12$ a.u. in Fig.~\ref{fig2}.

The ionization yield of direct process is computed in a similar way:

\begin{equation}
P_{\mathrm{dir}}=\int\left(\frac{d \sigma_{\mathrm{ion}}}{d p}\right) W(p) d p ,
\end{equation}

\noindent where the ionization cross section ${d\sigma_{\mathrm{ion}}}/{dp}$ is computed numerically as described in sec. \ref{sec:CS}. The procedures for computing $P_{\mathrm{dir}}$ are similar to the procedures used in calculating $P_{\text{RESI}} $ for both the cases of UU and UD ions. The dependence of $P_{\mathrm{dir}}$ on the field amplitude is shown in Fig. \ref{fig2} as ocher squares. The dynamics of direct ionization is essentially similar to that of RESI, its most striking property being the significantly lower magnitude in comparison with RESI which is explained mostly by the low values of direct ionization cross sections, as can be seen in Fig.~\ref{fig_cross}. As a result, direct ionization does not essentially contribute to the overall double ionization of the three-electron model.

As can be seen in Fig. \ref{fig2}, the yields of direct ionization as well as the RESI yield for the first D electron, that involves UU ion, experience a jump at certain values of $F$. Such a jump happens when the maximum first electron recollisional energy reaches the corresponding threshold value for excitation and ionization. The jump for RESI for the first U electron, that involves UD ion, is not pronounced because of the small magnitudes of laser-induced ionization rates from the first excited state for small fields (see Fig.~\ref{fig3}). The excitation cross sections from the ground state to excited states with higher energy are much smaller (see Fig.~\ref{fig_cross}) and thus play a lesser role in RESI even though the laser-induced ionization yields from these states are saturated over the whole range of field amplitudes in Fig.~\ref{fig3}. The differences between the magnitudes of the ionization channels are determined mainly by the threshold momentum/energy values, as well as by the values of total cross sections. The high momentum threshold value effectively cuts off the most intense part of the wavepacket, thus reducing the yield of a reaction. At the same time, the small total cross section leads to a decrease of the total reaction yields as well. In particular, the RESI yield of the UD model is drastically below that of the UU model for field values below 0.15 a.u. because of the much larger excitation energy for the main excited state and a smaller amplitude of the corresponding excitation cross section in the case of UD ion. The high values of ionization potentials, in comparison to the values of the corresponding excitation energies, together with the respective low values of ionization cross sections make the direct ionization yields always less intense than the yields of the corresponding RESI channels. This relation between RESI and direct ionization is in agreement with the estimations provided for realistic 3D atoms \cite{Chen2020-vw}.

\begin{figure}[t] 
	 \includegraphics[width=1.0\linewidth]{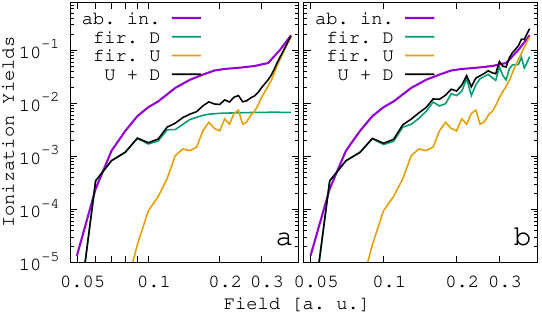}
	 \caption{ Double Ionization Yields in dependence on laser field amplitude. Violet curve correspond to full yield obtained with \textit{ab initio} three-electron model (sec. \ref{sec:3Emodel}). Green and ocher curves correspond to yield  after QRS calculations for the first ionized D electron and U electron (that is, UU and UD single ions formed as a result of such ionization correspondingly), while black curve is a simple sum of these two. In (a) the data from Fig.~\ref{fig2} has been used, in (b) the UU ion single ionization yield has been normalized with respect to first D electron yield after SFA calculations.  }
   \label{fig4}
\end{figure}

\subsection{Discussions of the results\label{sec:compar}}

The dependencies of the total double ionization yield $P_{\text{tot}}$ obtained with the ab initio model as well as with the semi-analytical model for both ionization scenarios are plotted in Fig.~\ref{fig4}~(a). The corresponding curves clearly indicate that the first D electron scenario is responsible for the early knee in the data obtained with the three-electron ab initio model. Despite the tunneling yield of the D electron being about two orders smaller than the corresponding yield of the upper U electron, the larger excitation cross sections together with significantly smaller excitation energies (see Fig.~\ref{fig_cross}) of the UU ion make RESI (UU) dominating over the double ionization channels for small fields. It forms a knee in the lower fields range that is continued to the higher fields by impacts of other ionization channels, notably sequential ionization of UU ion after the first escaped D electron and RESI involving the first U electron. Interestingly, the first U electron scenario never dominates in the knee regime, only the escape from the knee regime for $F>0.35$ a.u. is provided by sequential ionization after the first U electron.

The results of Fig.~\ref{fig4}~(a) have been obtained by imposing a saturation of a single ionization of the UU ion at the level of $10^{-2}$, since we assumed that this value is the upper limit of the laser-induced ionization yield of the first D electron. However, it is instructive to see how this assumption impacts the final results for double ionization yield. 
An alternative but still simple approach can be proposed to describe the SDI (UU) channel without restricting the value of tunneling ionization of a D electron in the first step. In such a case, we normalize the D electron yield after SFA by setting its magnitude for $F=0.07$ a.u. in correspondence with the SI (D) channel magnitude after three-electron ab initio calculations for the same value of $F$. For each field value, we multiply the numerically obtained single yield of the UU ion by the normalized SFA yield of D electron. In this way, the D-then-U double ionization channel intensity is not restricted from above. The results after the described correction can be seen in Fig.~\ref{fig4}~(b). The corresponding double ionization yield after the semi-analytic model describes the magnitude of the double ionization signal of the ab initio three-electron model significantly better than the corresponding curve in Fig.~\ref{fig4}~(a). Still, one can notice that even after this modification, the magnitude of the ab initio double ionization yield is up to 5 times larger than the corresponding curve for the semi-analytical model for the middle values of $F$. This discrepancy is probably caused by not accounting for all possible ionization channels in the semi-analytical model, as well as by the limited precision of SFA in predicting the magnitude of the ionization yield \cite{Chen2019-di}.

A striking observation can be made about the mismatch between the dynamics of particular double ionization channels defined within the three-electron ab initio model and those introduced in the semi-analytical model. For $F<0.30$ a.u., the numerical RII (recollisionally induced ionization) channel for D and U electrons dominates the competing channels, while QRS-based direct ionization channels have a negligible intensity compared to the other channels. The geometrical definition of RII \cite{Efimov2019} suggests that it accounts for electrons that escape the atom simultaneously or with a small time delay. Thus, if there is actually a nonessential delay between recollisionally induced excitation and laser-assisted ionization of the corresponding excited state, such a process can technically count as RII. This explains the domination of the RII(0-UD) yield in small fields. For $0.15<F<0.30$ a.u. sequential or quasi-simultaneous laser-assisted ionization of D and then U electrons can be numerically counted as RII as well because of the low time delay between the two events. The RII (0-UU) channel is suppressed by the spatial antisymmetry of the wavefunction with respect to the spatial coordinates of the two U electrons \cite{Efimov2019}, which virtually prohibits any near-simultaneous escape of the two electrons.

The correspondence between TDI channels of the ab initio model and the RESI and sequential ionization from the semi-analytical model is simpler. All events of the latter with some considerable time delay should constitute TDI signals. TDI (0-D-U) signal most probably accounts only for the first-D RESI (UU) events, since the D-then-U SDI corresponds to small time delays, as discussed earlier. This explains why the TDI (0-D-U) curve slightly decreases for $F>0.13$ a.u.: it generally follows the shape of RESI dependence in Fig.~\ref{fig2}~(a). The dominance of the TDI (0-U-D) and TDI (0-U-U) channels for $F>0.30$ is in perfect correspondence with the analogous behavior of the single ionization curve of the UD ion: after the first U electron escapes, it is followed by one of the electrons (U or D) that remain in the UD single ion.

At the same time, ab initio simulations show a relatively high magnitude of TDI (0-U-D) for $F<0.10$ a.u. This result is quite counterintuitive since for the given range of fields the RESI channel in the first-U scenario is more than one order weaker than RESI in the first-D scenario, according to the semi-analytic model. The available analytical tools seem to be incapable of providing an explanation of this numerical result. The results can become clear if one accepts the reasonable hypothesis that for lower field intensities RESI predominantly occurs through doubly excited states of a neutral atom, and not through excited states of a single ion\cite{Liao2017-uh}. In such a picture, the recolliding electron is captured by the parent ion, forming a doubly excited complex which subsequently decays into a system of a free electron and an excited single ion which can be further ionized by field. In such a case, the information about spin of the recolliding electron is lost, since the decay of a UD- doubly excited state can start with either a U or D electron with equal probability. This would explain very similar intensities of TDI (0-U-D) and TDI (0-D-U) channels within the indicated field range.

\section{Conclusions}

In this work, we have developed a combined semi-analytical approach based on QRS that contains a single-electron SFA model together with a set of two-electron ab initio models. It has allowed us to reproduce the dependence of the double ionization yield in the three-electron model on laser field intensity after the ab initio model with good precision. The semi-analytic model not only allows one to reduce the numerical complexity of simulating double ionization of the three-electron atom, but also, importantly, to identify particular channels that shape the different parts of the double ionization dependence. For development of the semi-analytic model, we created a version of Strong Field Approximation routines modified for simulating electronic escape and rescattering in a single-dimensional system. A set of ionization and excitation cross sections for a 1D system has been computed to support the QRS model.

The success of the semi-analytical model supports out initial assumptions of (i) competition of the two scenarios of ionization, when either U or, correspondingly, D electron is laser-ejected first and (ii) enhanced tunneling ionization of D orbital due to exchange interaction of bound electrons in an atom.
After them, double ionization yield of the three-electron atomic system is composed of the corresponding yields for the two scenarios. Those yields are, in turn, decomposed into yields of particular ionization channels: sequential, RESI and direct impact ionization.

We have found that the left part of the double ionization knee, which corresponds to lower field amplitudes, can be undoubtedly explained by the RESI occurring after the laser-induced escape of the D electron. This result is counterintuitive because of the very large ionization potential of D electron 1.21 a.u. The knee shape in the middle values of the external field is dominated by sequential ionization of the UU ion that resides after the ejection of the D electron. Only for large fields does the first-U scenario start to dominate the total double ionization through its sequential channel. There is still no clear understanding of the mechanism of sequential ionization in the first-D scenario. The corresponding ionization magnitude may vary in dependence on choice of a particular theoretical model.
Additional analysis allowed one to identify the probable reasons for the remaining discrepancy between the data after ab initio simulations with a three-electron model and the results after the developed semi-analytic model.

One can try to estimate expectations of how the double-ionization knee would be deformed in three-dimensional models if three-electron configurations are taken into account.
In single-dimensionality models the recollision intensity is overestimated, so in a full-dimensional model one would expect relatively smaller yield of RESI and direct channels. In such a case, the beginning of the knee in low field intensities would be diminished and additionally smoothed. At the same time, sequential ionization properties are supposed to have a weak dependence on the model dimensionality, and thus the amplified middle part of the knee should be preserved in the case of 3D atoms. Thus, the effect of three-electron configuration on double-ionization yield should be clearly visible in three-dimensional models as well.

\begin{acknowledgments}
The author thanks Jakub Prauzner-Bechcicki and Artur Maksymov for helpful discussions and Michał Ogryzek for his technical assistance with performing calculations. We gratefully acknowledge Polish high-performance computing infrastructure PLGrid (HPC Center: ACK Cyfronet AGH) for providing computer facilities and support within computational grant no. PLG/2025/018418.
\end{acknowledgments}


\bibliographystyle{apsrev4-2}
\bibliography{paper_bib}

\end{document}